\documentclass[final,5p,times,twocolumn]{elsarticle}
\usepackage{amssymb}
\journal{Physica C}

\begin{document}

\begin{frontmatter}

\title{Resonant inelastic x-ray scattering in single-crystal superconducting PrFeAsO$_{0.7}$}

\author[1,2]{I. Jarrige} 
\author[1,2]{K. Ishii} 
\author[1,3]{M. Yoshida}
\author[1,2,4]{T. Fukuda}
\author[1]{K. Ikeuchi \fnref{fn1}}
\author[2,5]{M. Ishikado}
\author[6]{N. Hiraoka}
\author[6]{K.D. Tsuei}
\author[2,7]{H. Kito}
\author[2,7]{A. Iyo}
\author[2,7]{H. Eisaki}
\author[2,5]{S. Shamoto}
\address[1]{Synchrotron Radiation Research Center, Japan Atomic Energy Agency, Hyogo 679-5148, Japan}
\address[2]{JST, TRIP, 5, Sanbancho, Chiyoda, Tokyo 102-0075, Japan}
\address[3]{Department of Physics, Tohoku University, Sendai, 980-8578, Japan}
\address[4]{Materials Dynamics Laboratory, SPring-8/RIKEN, Sayo, Hyogo 679-5148, Japan}
\address[5]{Quantum Beam Science Directorate, Japan Atomic Energy Agency, Tokai, Ibaraki 319-1195, Japan}
\address[6]{National Synchrotron Radiation Research Center, Hsinchu Science Park, Hsinchu 30076, Taiwan}
\address[7]{Nanoelectronics Research Institute (NeRI), National Institute of Advanced Industrial Science and Technology (AIST), 1-1-1 Central 2, Umezono, Tsukuba, Ibaraki, 305-8568, Japan}
\fntext[fn1]{Present address: Institute of Materials Structure Science, High Energy Accelerator Research Organization (KEK), Tsukuba 305-0801, Japan}

\begin{abstract}
Resonant inelastic x-ray scattering (RIXS) spectra at the Fe $K$-edge were measured for a single crystal of the iron oxypnictide superconductor PrFeAsO$_{0.7}$ ($T_{c}$=42 K). They disclose a weak, broad feature centered around 4.5~eV energy loss, which is slightly resonantly enhanced when the incident energy is tuned in the vicinity of the $4p$ white line. We tentatively ascribe it to the charge-transfer excitation between As~$4p$ and Fe~$3d$.
\end{abstract}

\begin{keyword}
Iron pnictides \sep Resonant inelastic x-ray scattering
\end{keyword}
\end{frontmatter}

\section{Introduction}

The recent discovery of high-Tc superconductivity in the iron oxypnictides \cite{kamihara} was shortly followed by the attainment of the highest critical temperature in all non-cuprate materials for the samarium-based system \cite{ren}. It was rapidly ensued by a plethora of experimental and theoretical works on the electronic properties of this new class of compounds. Interestingly enough, their electronic structure is not fully elucidated yet. It is usually described by one or the other of two theoretical approaches, the first one suggesting an itinerant-electron state \cite{vildosola,anisimov}, and the second one underlining on-site correlations \cite{haule,craco}. Seemingly discordant results were obtained experimentally as well, with angle-resolved photoelectron spectroscopy favoring an itinerant electronic behavior \cite{lu} and optical conductivity pointing to a somewhat correlated electronic structure \cite{boris}. This fundamental cleavage highlights the need for more compelling insight into the electronic structure of the iron-based superconductors which, we believe, could be provided by resonant inelastic x-ray scattering (RIXS).

RIXS has won recognition over the past decade as a potent tool for characterization of the low-lying electronic excitations, owing to i) the resonant enhancement of the scattering cross section near the core-excitation threshold, ii) momentum resolution, and iii) bulk sensitivity. Whilst RIXS can offer a sound diagnosis of the electronic properties of the cuprates \cite{hill,ishii}, its suitability to the iron-based superconductors could be put to question because of the more delocalized character of the Fe $d$ bands. Indeed, in RIXS, charge excitations are mostly brought about by the Coulomb interaction between the core-hole potential and the valence electrons in the intermediate state. The more delocalized the valence electrons are, the weaker their interaction with the core hole gets. Notwithstanding these difficulties, we here report on an examination of the electronic structure of a single-crystal of superconducting PrFeAsO$_{0.7}$ employing RIXS at the Fe $K$-edge. We discern a broad feature near 4.5~eV, reminiscent of a ligand-to-metal charge-transfer excitation.

\section{Experimental details}

The data were taken at the Taiwan beamline BL12XU at SPring-8 with both the horizontal scattering plane and the incident photon polarization ($\varepsilon$) parallel to the $ac$ plane. A total energy resolution of 1 eV was achieved using a double-crystal monochromator Si (111) and a 1-m bent Ge (620) crystal analyzer for the incident and scattered beams, respectively. The size of the beam at the sample position was 24~$\mu$m (horizontal) x 37~$\mu$m (vertical). A single crystal of PrFeAsO$_{0.7}$ was grown by a high-pressure synthesis method using a belt-type anvil apparatus \cite{ishikado}. 

\section{Results and discussion}

\begin{figure}[!ht]
\includegraphics [width=8cm] {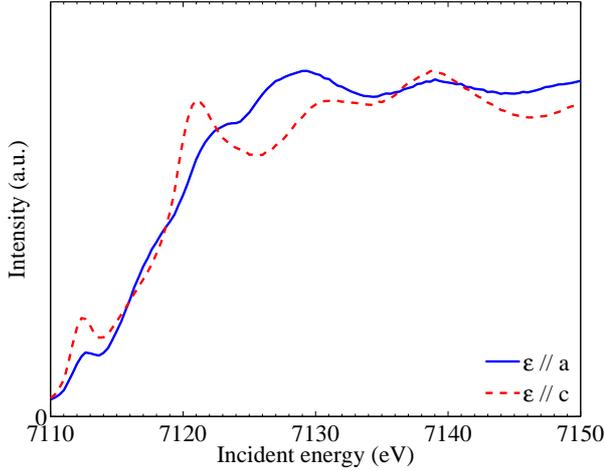}
\caption{Absorption spectrum measured at the Fe $K$-edge on PrFeAsO$_{0.7}$ for both $\varepsilon // a$ and $\varepsilon // c$ polarization conditions.}
\end{figure}

The absorption spectrum measured in the total fluorescence yield mode at the Fe $K$-edge for incident photon polarization $\varepsilon$ both parallel ($\varepsilon // a$) and perpendicular ($\varepsilon // c$) to the FeAs planes is shown in Fig. 1. The maximum of the white line, corresponding to transitions to the Fe $4p$ states making antibonding interactions with As electronic orbitals, is shifted towards the high-energy side in the $\varepsilon // a$ geometry (7129 eV) compared with $\varepsilon // c$ (7121 eV). This is concordant with the fact that, in the former geometry, one probes the Fe orbitals that lie within the FeAs plane and interact most with the As orbitals, hence the shift of the antibonding states towards high energies. For the same reason, the pre-edge, mostly relating to the Fe $3d$ states, is seen to dwindle and slightly broaden for $\varepsilon // a$ compared with $\varepsilon // c$.

The incident-energy (E$_{1}$) dependence of the RIXS spectrum was measured across the 7110-7130 eV range at the (0,0,6.5) tetragonal reciprocal lattice point. The corresponding scattering angle is approximately 83$^o$, where a relatively reduced elastic scattering can be achieved. Apart from the prominent $K\beta _{5}$ fluorescence peak that drifts linearly with E$_{1}$, almost featureless spectra were obtained, at stark contrast with the usual RIXS signal obtained on the cuprates. This flagrant difference may stem from the itinerant nature of the Fe $3d$ electrons in the iron oxypnictides, resulting in a large bandwidth, and therefore a weak interaction with the core-hole potential in the RIXS intermediate state. Another plausible reason could be that the $K\beta _{5}$ fluorescence line, because the Fe site is not centrosymmetric, gains intensity from both quadrupolar ($3d \rightarrow 1s$) and dipolar ($4p$-$3d \rightarrow 1s$) transitions, the latter occuring through $4p$-$3d$ hybridization. This behavior would be analogous to that in the pre-edge of the $K$ absorption spectra of $3d$ metals \cite{rueff}. A stronger intensity would accordingly ensue for the $K\beta _{5}$ of Fe in the oxypnictides compared with the cuprates where Cu occupies a centrosymmetric site and therefore yields a pure quadrupolar $K\beta _{5}$ line.

\begin{figure}[!ht]
\includegraphics [width=8cm] {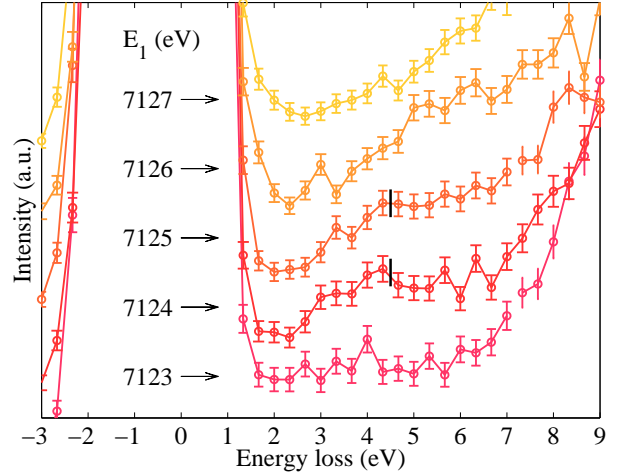}
\caption{Incident-energy dependence of the RIXS spectrum of PrFeAsO$_{0.7}$. The strong tail on the high-energy side is the onset of the $K\beta _{5}$ fluorescence line.}
\end{figure}

Our main finding is a faint, broad peak centered at $\sim$4.5~eV which undergoes a slight resonant enhancement at constant loss energy when E$_{1}$ is set to 7124$\sim$7125~eV (cf. vertical ticks in Fig. 2). The energy of this feature is reminiscent of the charge transfer from O~$2p$ to Cu~$3d$ in the cuprates, it is therefore tempting to assign it to the charge transfer between As~$4p$ and Fe~$3d$. According to first-principle calculations of the local densities of states (DOS) of LaFeAsO by Ishibashi $\textit{et al.}$ \cite{ishibashi}, the maximum of the As~$4p$ occupied states lies around - 3 eV and the Fe~$3d$ unoccupied states show a peak near 1~eV. We suggest therefore to assign the 4.5-eV RIXS feature to the charge transfer between As~$4p$ and Fe~$3d$. We plan to refine our analysis via further measurements with a higher energy resolution.

\section{Summary}

We have studied the incident-energy dependence of the RIXS spectrum of the superconductor PrFeAsO$_{0.7}$. The spectra are found to be rather featureless, which could be explained either by the itinerant nature of the Fe $3d$ electrons, or by the strong intensity of the $K\beta _{5}$ line that could hide some of the RIXS features, or by both. The only feature which we observed is a broad peak near 4.5-eV energy loss, lying in between the elastic peak and the tail of the $K\beta _{5}$ fluorescence line. Based on a comparison with DOS calculations, this feature is likely related to the charge transfer excitation between As~$4p$ and Fe~$3d$.

\end{document}